\documentclass{emulateapj}
\usepackage{apjfonts}
\lefthead{Kim et al.} 
\righthead{PIXEL-LENSING DETECTIONS OF M31 BINARIES}

\begin{document}
\title{Detection of M31 Binaries via High-Cadence Pixel-Lensing Surveys}

\author{
D. Kim\altaffilmark{1},
S.-J. Chung\altaffilmark{1}\\
and\\
M.\ J. Darnley\altaffilmark{2},
J.\ P. Duke\altaffilmark{2},
A. Gould\altaffilmark{5},
C. Han\altaffilmark{1, 8},
M.\ A. Ibrahimov\altaffilmark{7},
M. Im\altaffilmark{6},
Y.-B. Jeon\altaffilmark{4},
R.\ G. Karimov\altaffilmark{7},
E. Kerins\altaffilmark{2, 3},
C.-U. Lee\altaffilmark{4},
A. Newsam\altaffilmark{2}, and
B.-G. Park\altaffilmark{4}\\
(The Angstrom Collaboration)
}

\altaffiltext{1}{Department of Physics, Institute for Basic Science
Research, Chungbuk National University, Chongju 361-763, Korea}
\altaffiltext{2}{Astrophysics Research Institute, Liverpool John Moores
University, Twelve Quays House, Birkenhead, Merseyside CH41 1LD, United Kingdom}
\altaffiltext{3}{Jordell Bank Observatory, University of Manchester,
Macclesfield, Cheshire SK11 9DL, United Kingdom}
\altaffiltext{4}{Korea Astronomy and Space Science Institute, Hwaam-Dong,
Yuseong-Gu, Daejeon 305-348, Korea}
\altaffiltext{5}{Department of Astronomy, The Ohio State University,
140 West 18th Avenue, Columbus, OH 43210}
\altaffiltext{6}{Department of Physics and Astronomy, 
Seoul National University, Seoul 151-742, Korea}
\altaffiltext{7}{Ulugh Beg Astronomical Institute, Uzbek Academy of Sciences, 
Tashkent, Uzbekistan}
\altaffiltext{8}{Corresponding author}



\begin{abstract}
The Angstrom Project is using a distributed network of two-meter class 
telescopes to conduct a high cadence pixel-lensing survey of the bulge of 
the Andromeda Galaxy (M31).  With the expansion of global telescope network, 
the detection efficiency of pixel-lensing surveys is rapidly improving.
In this paper, we estimate the detection rate of binary lens events 
expected from high-cadence pixel-lensing surveys toward M31 such as the 
Angstrom Project based on detailed simulation of events and application 
of realistic observational conditions.  Under the conservative detection 
criteria that only high signal-to-noise caustic-crossing events with long 
enough durations between caustic crossings can be firmly identified as 
binary lens events, we estimate that the rate would be $\Gamma_{\rm b}\sim
(7-15)f_{\rm b}(N/50)$ per season, where $f_{\rm b}$ is the fraction of 
binaries with projected separations of $10^{-3}\ {\rm AU}<\tilde{d}<10^3\ 
{\rm AU}$ out of all lenses and $N$ is the rate of stellar pixel-lensing 
events.  We find that detected binaries would have mass ratios distributed 
over a wide range of $q\gtrsim 0.1$ but with separations populated within 
a narrow range of $1\ {\rm AU}\lesssim \tilde{d}\lesssim 5\ {\rm AU}$.  
Implementation of an alert system and subsequent follow-up observations 
would be important not only for the increase of the binary lens event rate 
but also for the characterization of lens matter.
\end{abstract}


\section{Introduction}

Surveys to detect transient variations of stellar brightness caused by 
gravitation microlensing have been and are being conducted towards 
various star fields.  These fields include the Magellanic Clouds (MACHO: 
\citet{alcock00}, EROS: \citet{afonso03}), Galactic bulge (MACHO: 
\citet{alcock01}, EROS: \citet{hamadache06}, OGLE: \citet{sumi06}, 
MOA: \citet{bond01}), and M31 (AGAPE: \citet{ansari99}, POINT-AGAPE: 
\citet{calchinovati05}, VATT-Colombia: \citet{uglesich04}, MEGA: 
\citet{dejong04}, WeCAPP: \citet{riffeser03}).  The number of lensing 
events detected so far from these surveys is about 3000.  Most of these 
events were detected toward the Galactic bulge field and the number of 
events detected toward the Magellanic Cloud and M31 fields, which are 
several dozens toward the individual fields, have meager contributions 
to the total number of events.  Towards the Magellanic Cloud field, the 
line of sight passes mainly through the halo of our Galaxy and thus the 
low detection rate is attributed to the scarcity of massive compact halo 
objects that can work as lenses \citep{alcock00, tisserand06}.  On the 
other hand, the line of sight toward M31 passes through the dense stellar 
region of M31 and thus the small number of detected stellar lensing events 
is mainly due to the low detection efficiency of the surveys.

With the expansion of global telescope network, however, the detection 
efficiency of M31 lensing surveys is expected to greatly improve.  For 
example, a new M31 pixel-lensing survey, the Andromeda Galaxy Stellar 
Robotic Microlensing (Angstrom) 
Project\footnote{{\sf http://www.astro.livjm.ac.uk/angstrom/}}
will be able to achieve a monitoring frequency of $\sim 5$ observations 
per 24-hour period by using a network of telescopes, including the robotic 
2m Liverpool Telescope at La Palma, Faulkes Telescope North in Hawaii, 
1.8m telescope at the Bohyunsan Observatory in Korea, the 2.4m Hiltner 
Telescope at the MDM Observatory in Arizona, and 1.5m at the Maidanak 
Observatory in Uzbekistan.  Intensive monitoring programs such as 
Angstrom are expected to detect events with a rate of $\sim 100$ per 
season \citep{kerins06}.  Among them, a considerable fraction would be 
caused by binaries.  \citet{baltz01} pointed out that the rate of binary 
events relative to single lens events is higher in pixel-lensing because 
many of binary events involve caustic crossings during which the 
magnification is high and thus more detectable.  In this work, we estimate 
the detection rate of binary lens events expected from a high cadence 
pixel-lensing survey toward M31 based on detailed simulation of events 
and application of realistic observational conditions.

The paper is organized as follows.  In \S\ 2, we describe basics of binary 
lensing.  In \S\ 3, we investigate the dependency of the detection efficiency 
on the binary lens parameters.  In \S\ 4, we estimate the event rate based 
on the simulation of M31 pixel-lensing events.  We describe the details of 
the simulation and the criteria applied for the selection of binary events.
We also investigate the characteristics of the binaries detectable from
the survey.  We end with a brief summary of the results and discussion in 
\S\ 5.

\begin{figure*}[t]
\epsscale{0.7}
\plotone{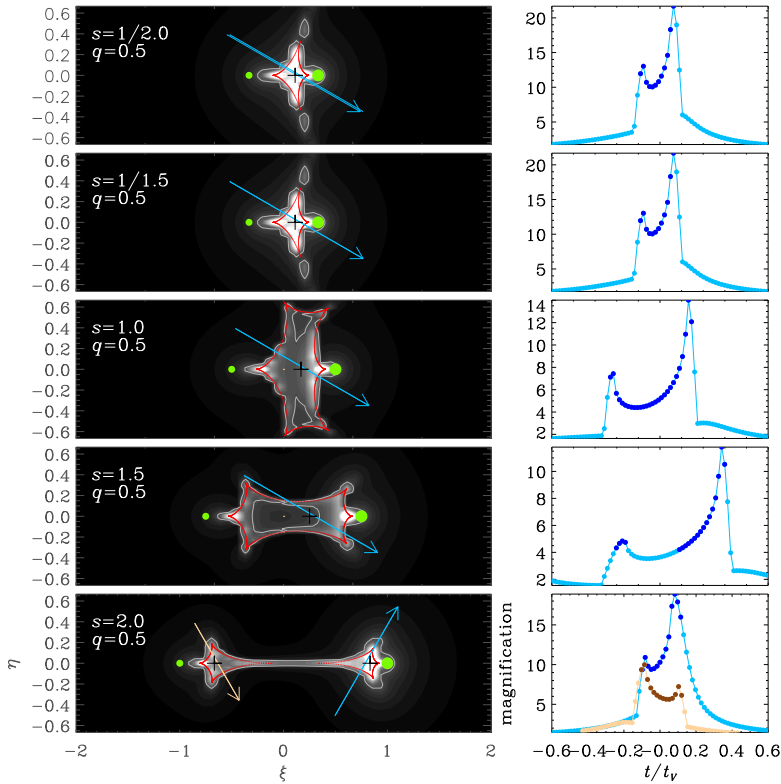}
\caption{\label{fig:one}
Variation of caustic pattern depending on binary separation and example 
light curves of binary lens events.  In each of the left panels, the green 
dots represent the positions of the individual binary lens components, 
where the heavier lens is toward right.  The coordinates are centered at 
the midpoint between the binary lens components.  The figure drawn by red 
curve represents the caustic.  The cross mark denotes the effective 
single lens position, which roughly corresponds to the center of each 
set of caustics.  The greyscale represents the signal-to-noise ratio of 
lensing-induced source flux variation as a function of source position.  
A contour (white curve) is drawn at the level of $S/N=3.0$.  Presented in 
the right panels are the light curves of binary lens events resulting from 
the source trajectories marked by arrow in the corresponding left panels.  
For the case where $s=2.0$, we present two light curves and the source 
trajectories responsible for the individual light curves are marked by 
the corresponding colors.  The dark-tone dots on each light curve represent 
the data points satisfying the condition that the source is within the 
caustic and the signal-to-noise ratio is higher than a threshold value of 
$(S/N)_{\rm th} =3.0$.
}\end{figure*}

\section{Basics of Binary Lensing}

General relativity predicts that a light ray passing by a stellar object
is deflected.  If a source star is gravitationally lensed by a binary 
lens, the equation of lens mapping from the lens plane to the source plane 
is expressed as
\begin{equation}
\zeta = z - \sum_{j=1}^2 {m_j/M \over \bar{z}-\bar{z}_{L,j}},
\label{eq1}
\end{equation}
where $\zeta=\xi + i\eta$, $z_{{\rm L},j}=x_{{\rm L},j}+iy_{{\rm L},j}$, 
and $z=x+iy$ are the complex notations of the source, lens, and image 
positions, respectively, $\bar{z}$ denotes the complex conjugate of $z$, 
$m_j$ are the masses of the individual lens components, and $M=m_1+m_2$ 
is the total mass of the system \citep{witt90}.  Here all lengths are 
normalized to the radius of the Einstein ring of the total mass of the 
system.  The Einstein radius is related to the physical parameters of the 
lens system by
\begin{equation}
\theta_{\rm E}=
\left( {4GM\over c^2} \right)^{1/2}
\left( {1\over D_{\rm L}} - {1\over D_{\rm S}}  \right)^{1/2},
\label{eq2}
\end{equation}  
where $D_{\rm L}$ and $D_{\rm S}$ are the distances to the lens and source, 
respectively.  Due to lensing, the source star image is split into several 
segments.  For binary lensing, the number of images, $N_I$, is either three 
or five depending on the source position with respect to the lens position.
The lensing process conserves the source surface brightness and thus the 
magnifications $A_i$ of the individual images correspond to the ratios 
between the areas of the images and source.  For an infinitesimally small 
source element, the magnification is
\begin{equation}
A_i = \left\vert \left( 1-{\partial\zeta\over\partial\bar{z}}
{\overline{\partial\zeta}\over\partial\bar{z}} \right)^{-1} \right\vert.
\label{eq3}
\end{equation}
Then, the total magnification corresponds to the sum over all images, 
$A=\sum_i^{N_I} A_i$.

The main new feature of binary lensing compared to single lensing is the 
formation of caustics.  Caustics are the set of positions in the source 
plane on which the magnification of a point-source event is infinite.  
The set of caustics form closed curves, which are composed of multiple 
concave line segments (fold) that meet at points (cusp).  When the source 
enters the caustic curve, two new images appear and the number of images 
changes from three into five.  Due to the divergent nature of the 
magnification near a caustic, the light curve during the caustic crossing
of a binary lens event is characterized by a sharp spike.  Since the 
caustic curve is closed, the number of caustic crossings is a multiple 
of two.  The light curve between a set of two caustic crossings is 
characterized by its distinctive ``U''-shape.

The number and shape of caustics vary depending on the separation, $s$ 
(normalized by $\theta_{\rm E}$), and mass ratio, $q$, between the two 
lens components.  If the separation is substantially smaller than the 
Einstein radius, $s\ll 1$, there exist three sets of caustics.  One big 
caustic is located close to the center of mass of the binary and the 
other two tiny ones are located away from the center of mass on the 
heavier lens side.  If the separation is substantially larger than the 
Einstein radius, $s\gg 1$, on the other hand, there exist two sets 
of caustics.  They are located close to the positions of the individual 
lens components but slightly shifted toward the lens component of the 
other side.  The amount of the shift is 
\begin{equation}
\Delta z_{{\rm L},i} = {\rm sgn}(z_{{\rm L},j}-z_{{\rm L},i})
{m_j/M \over s},
\label{eq4}
\end{equation}
where $z_{{\rm L},i}$ is the position of the lens component on the same 
side of each set of caustics while $m_j/M$ and $z_{{\rm L},j}$ represent 
the mass fraction and position of the other lens component located on the 
opposite side, respectively \citep{distefano96}.  If the binary separation 
is equivalent to the Einstein ring, $s\sim 1.0$, the caustic curve forms 
a single large closed figure with its center roughly at the center of mass 
of the binary.  In Figure~\ref{fig:one}, we present the caustic pattern 
for various binary lenses with different separations.  The dependence of 
the caustic size on the mass ratio is weak.  As a result, the caustic size 
is not negligible even for an extreme case of a star-planet binary lens 
pair with $q\lesssim 10^{-3}$.  This makes microlensing an efficient 
method for planet searches \citep{mao91}.

\section{Detection Efficiency}

Due to the variation of the caustic characteristics depending on the 
binary lens parameters, the detection efficiency of binary lens events 
also depends on these parameters.  In the section, we investigate the
dependencies of the detection efficiency on the binary lens parameters.

We estimate the efficiency for an example M31 event.  We choose a 
representative event based on the simulation of M31 pixel-lensing events 
conducted by \citet{kerins06}.  In the simulation, M31 matter density 
distribution is based on standard double-exponential disk plus triaxial 
bulge models.  The velocity is modeled by an isotropic Maxwellian 
distribution.  The mass function of lenses are modeled by a broken 
power-law represented by
\begin{equation}
\Phi(m) = 
\cases{
\kappa(m/0.5\ M_\odot)^{-1.4},  & for $0.08\ M_\odot \leq m < 0.5\ M_\odot$, \cr
\kappa(m/0.5\ M_\odot)^{-2.35}, & for $0.5\ M_\odot \leq m < 1.0\ M_\odot$, \cr
}
\label{eq5}
\end{equation}
where $\kappa$ is a proportional constant.  According to this simulation, 
the total number of events expected to be detected by the Angstrom 
survey is $\sim 30-60$ events per season over an area of $11\times 11\ 
{\rm arcmin}^2$ of the M31 bulge field under the assumption that all 
lenses are composed of a single stellar component.  Based on these events, 
we select a representative event as the one caused by a binary lens system 
with a primary lens mass of $m=0.4\ M_\odot$.  The assumed distances to 
the source star and lens are $D_{\rm S} =780$ kpc and $D_{\rm L}=(780-0.8)$ 
kpc, respectively.  For a representative value of the Einstein timescale 
(time needed for the source to transit the Einstein radius of the primary 
lens), we adopt $t_{\rm E}=10$ days based on the simulation of 
\citet{kerins06}.  For the source star, we choose a giant with a brightness 
$I\sim 24.4$.  We consider finite-source effects by adopting a source radius 
of $R_\star=10.0\ R_\odot$.  We assume an $I$-band background brightness of 
$\mu_I =17.6\ {\rm mag}/ {\rm arcsec}^2$.  We let the mass ratio and 
separation between binary lens components vary as free parameters.

Once the lensing parameters are set, we then produce lensing events.  
Due to the high background flux level, detections of lensing events 
toward M31 field are limited only to those with high magnifications.  
In the background-dominated regime such as the M31 field, the 
signal-to-noise ratio of a pixel-lensing event is 
\begin{equation}
S/N = {F_{\rm S}t_{\rm exp}^{1/2}(A-1) \over F_{\rm B}^{1/2}},
\label{eq6}
\end{equation}
where $F_{\rm S}$ and $F_{\rm B}$ represent the flux from the source and 
background, respectively, and $t_{\rm exp}$ represents the exposure time.  
If the threshold signal-to-noise ratio is $(S/N)_{\rm th}$, the threshold 
magnification and the corresponding normalized (by $\theta_{\rm E}$) 
impact parameter of the source trajectory for event detection are defined 
respectively as
\begin{equation}
A_{\rm th} = 1+{(S/N)_{\rm th} F_{\rm B}^{1/2} \over F_{\rm S}}, 
\label{eq7}
\end{equation}
and
\begin{equation}
u_{0,{\rm th}}=\left[{2 \over (1.0-A_{\rm th}^{-2})^{1/2}}-2 \right]^{1/2}. 
\label{eq8}
\end{equation}
We, therefore, produce events with impact parameters of the source 
trajectory with respect to the {\it effective single lens position}, 
$z_{\rm L,eff}$, less than the threshold value defined in 
equation~(\ref{eq8}) with an adopted threshold signal-to-noise ratio of 
$(S/N)_{\rm th}=3.0$, i.e.\ $3\sigma$ detection.  The effective single 
lens position represents the location of a single lens at which the 
resulting single lensing light curve best describes the light curve of a 
binary lens event.  For example, the light curve of an event caused by a 
close binary lens is well described by the light curve of a single lens 
event caused by a mass equal to the total mass of the binary located at 
the center of mass of the binary.  In this case, the effective single 
lens position is the center of mass of the binary.  For the case of a 
wide-separation binary, on the other hand, the individual lens components 
behave as if they are two independent single lenses located roughly at 
the centers of the individual sets of caustics, i.e.\ $z_{{\rm L},{\rm eff}}
=z_{{\rm L}}+\Delta z_{{\rm L}}$.  In this case, there are two effective 
lens positions located at the center of the individual 
caustics.\footnote{We distinguish wide-separation binaries from close 
binaries if the caustic is divided into two pieces.}  In 
Figure~\ref{fig:one}, the effective single lens positions are denoted by 
cross marks for various cases of binary lenses.  Once events are produced, 
we estimate the detection efficiency of binary lens events as the ratio of 
the number of events that can be firmly identified to be caused by binary 
lenses out of all produced events.

Light curves of events are produced under the following observational
conditions.  For the photometry, we assume that the instrument can detect 
1 photon per second for an $I=24.2$ star following the specification of 
the Liverpool Telescope.  The observation is assumed to be carried out 
such that small-exposure images are combined to make a 30 min exposure 
image to obtain a high signal-to-noise ratio while preventing saturation 
in the central bulge region.  We assume that 5 such combined images are 
obtained per 24-hour period from the observations by using  network 
telescopes.  The photometry is done based on difference imaging 
\citep{alard98}.  The flux variation is assumed to be measured at an 
aperture that maximizes the signal-to-noise ratio of the measured flux 
variation.  Under the assumption of a Gaussian point spread function, 
the optimal aperture that maximize the signal-to-noise ratio is 
$\theta_{\rm ap}=0.673\theta_{\rm see}$ \citep{kerins06}.  We assume that 
the average seeing is $\theta_{\rm see}= 1''\hskip-2pt .0$.  With the 
adoption of this aperture, the fraction of the source flux within the 
aperture is $F(\theta\leq \theta_{\rm ap})/F_{\rm tot}=0.715$, where 
$F_{\rm tot}$ is the flux measured at $\theta_{\rm ap}\equiv \infty$.

\begin{figure}[t]
\epsscale{1.15}
\plotone{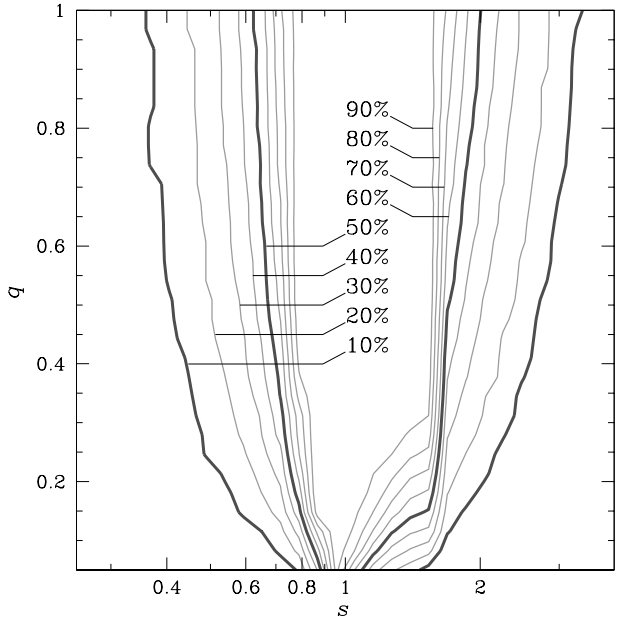}
\caption{\label{fig:two}
Detection efficiency of binary events toward the M31 bulge field as a 
function of the binary separation and mass ratio.  The efficiency is for 
a typical pixel-lensing event expected to be detected by a high-cadence 
pixel-lensing survey such as Angstrom.  For details about the representative 
event and observational conditions, see \S\ 3.
}\end{figure}

We assume the following criteria for the identification of binary lens 
events.  The light curves of binary events exhibit diverse patterns.  In 
many cases, it is difficult to firmly distinguish light curves from those 
caused by single lenses or other types of source variability.  Fortunately, 
a fraction of binary lens events involve caustic crossings and the 
characteristic features (e.g.\ spikes during caustic crossings and U-shape 
curve between caustic crossings) in the resulting light curves can be used 
to securely identify not only the lensing-induced variability but also the 
binary nature of lenses.  We, therefore, restrict detectable binary events 
only to caustic-crossing events.  For the confirmation of caustic crossings, 
we require events should have at least 5 data points with $S/N\geq 3.0$ on 
the part of the light curve between caustic crossings.  Considering that 
the assumed observation frequency is 5 times per day, this requirement is 
equivalent that source star should stay in the caustic at least for a day. 
We note that the data points with $S/N\geq 3.0$ do not have to be consecutive.

In Figure~\ref{fig:two}, we present the detection efficiency of binary 
pixel-lensing events toward the M31 bulge field as a function of the 
binary separation and mass ratio.  From the figure, one finds that binaries 
with separations $0.4 \lesssim s \lesssim 3.0$ can be efficiently detected.
Especially for binaries with separations very close to the Einstein radius, 
the efficiency is nearly 100\%.  One also finds that the dependency of the 
efficiency on the mass ratio is weak and thus detections can be extended 
to low-mass companions.

\section{Binary-Lens Event Rate}

Pixel-lensing events have diverse physical parameters.  For the estimation 
of the binary event rate, then, convolution of the efficiency estimated 
for a specific event with distributions of the physical lens parameters 
is required.

\begin{figure}[t]
\epsscale{1.20}
\plotone{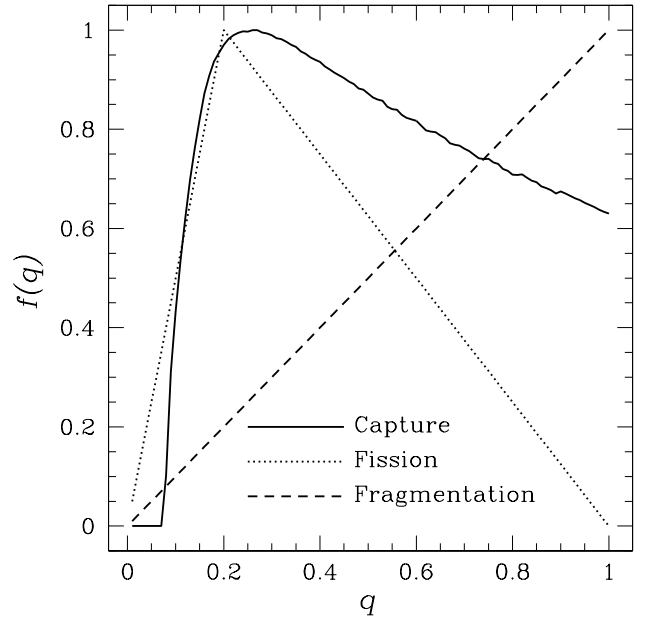}
\caption{\label{fig:three}
Models of binary mass-ratio distribution.
}\end{figure}

The physical parameters involved with primary lens components are based 
on the simulation of \citet{kerins06} described in \S\ 3.  Once these 
parameters are set, we then set the parameters related to companions.
Following \citet{abt83}, we model the binary separation as uniformly 
distributed in $\log s$, i.e.\ $f(\log s)\equiv {\rm const}$.  For the 
mass ratio distribution, we test three models.  The first model is based 
on the assumption that the two masses of the binary components are drawn 
independently from the same mass function as that of single stars.  This 
model is the natural result of binary formation process where binaries are 
formed through interactions between protostellar disks \citep{pringle89} 
or some other form of capture.  We refer to this model as the `capture' 
model.  Other possible mechanisms of binary formation are fission of 
a single star and fragmentation of a collapsing object.  Numerical 
calculations suggest that the former process results in a mass ratio 
distribution peaking at around $q=0.2$ \citep{lucy77}, while the latter 
results in more equal masses \citep{norman78}.  We refer to these models 
as the `fission' and `fragmentation' models, respectively.  We model the 
mass ratio distribution in the fission model as 
\begin{equation}
f(q)=
\cases{
5q           &  for $q\leq 0.2$, \cr
-1.25q+1.25  &  for $q>0.2$.    \cr
}
\label{eq9}
\end{equation}
The distribution of the fragmentation model is modeled as
\begin{equation}
f(q) = q.
\label{eq10}
\end{equation}
In Figure~\ref{fig:three}, we present the mass ratio distributions of the 
three tested models.  In the capture and fragmentation models, the mass of 
the primary is drawn from the single lens mass function and the mass of 
the companion is determined based on the mass ratio derived from the mass 
ratio distribution.  In the fission model, on the other hand, we set the 
total mass of the binary to be the same as that of a single lens because 
a single mass is split into two components of a binary in this model.  
Then, the total mass of the binary is drawn from the mass function of 
single lenses and the masses of the individual binary components are 
deduced from the mass ratio.  As a result, the average mass of the binary 
lenses in the fission model is smaller than those in the capture and 
fragmentation models.

\begin{figure}[t]
\epsscale{1.20}
\plotone{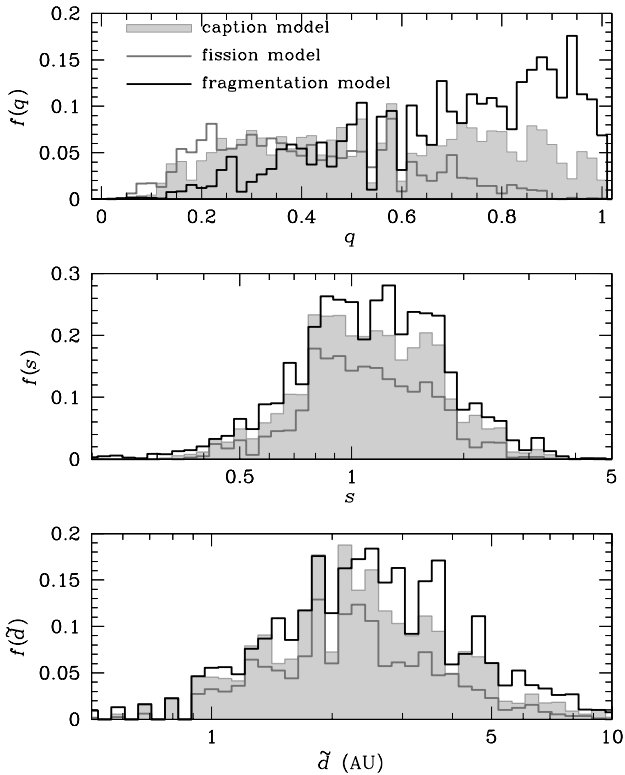}
\caption{\label{fig:four}
Distributions of mass ratios (upper panel), normalized separations 
(middle panel), and physical separations (bottom panel) of binary lenses 
expected to be detected by a high-cadence pixel-lensing survey towards M31.  
The three histograms in each panel represent the distributions under 
different models of binary mass-ratio distributions.
}\end{figure}

\begin{deluxetable}{lc}
\tablecaption{Rate of Binary Lens Events\label{table:one}}
\tablewidth{0pt}
\tablehead{
\colhead{mass ratio model} &
\colhead{binary event rate} 
}
\startdata
capture model        &  12.0 $f_{\rm b}$ events per season \\
fission model        &   7.4 $f_{\rm b}$ events per season\\
fragmentation model  &  15.2 $f_{\rm b}$ events per season
\enddata 
\tablecomments{ 
Rates of binary lens events expected to be detected by a high-cadence 
pixel-lensing survey toward M31 field under three different models of 
binary mass-ratio distributions.  The quantity $f_{\rm b}$ represents 
the fraction of binaries with projected separations of $10^{-3}\ {\rm AU}
<\tilde{d}<10^3\ {\rm AU}$ out of all lenses.}
\end{deluxetable}

In Table~\ref{table:one}, we present the expected rates of binary lens 
events under three different models of binary mass ratio distribution.  
We note that the event rate is normalized so that it becomes 50 events 
per season if all lenses are composed of a single component following 
the estimation of \citet{kerins06}.  We find that the rate of binary 
lens events is $\Gamma_{\rm b}\sim 7f_{\rm b}$ -- $15f_{\rm b}$ events 
per season, where $f_{\rm b}$ is the fraction of binary lenses with 
projected separations of $10^{-3}\ {\rm AU}<\tilde{d} <10^3\ {\rm AU}$ 
out of all lenses.  By adopting a value of $f_{\rm b}=0.5$ and considering 
the $\sim7$-month duration of M31 season (from August through to February), 
this rate roughly corresponds to $\sim 0.7$ events per month.  We find 
that the dependence of the detection rate on the binary mass ratio 
distribution is such that the rate increases as more companions are 
populated in higher mass-ratio region.  As a result, the rate becomes 
bigger in the order of the  fission, capture, and fragmentation models.
In Figure~\ref{fig:four}, we also present the distributions of mass ratios 
and separations (both normalized and physical separations) between the 
components of binary lenses.  From the figure, we find that binaries to 
be detected by high-cadence pixel-lensing surveys will have mass ratios 
distributed over a wide range of $q \gtrsim 0.1$.  By contrast, separations 
of binaries will be populated within a narrow range of $1\ {\rm AU}\lesssim 
\tilde{d}\lesssim 5\ {\rm AU}$.

\section{Discussion and Conclusion}

We estimated the detection rate of binary lens events expected from 
high-cadence pixel-lensing surveys toward M31 such as the Angstrom 
Project based on detailed simulation of events and application of 
realistic observational conditions.  Under the conservative detection 
criteria that only high signal-to-noise caustic-crossing events with 
long enough durations between caustic crossings can be firmly identified 
as binary lens events, we estimated that the rate would be $\Gamma_{\rm b}
\sim (7-15)f_{\rm b}(N/50)$ per season, where $N$ is the rate of stellar 
pixel-lensing events.  We found that detected binaries would have mass 
ratios distributed over a wide range of $q\gtrsim 0.1$ but with separations 
populated within a narrow range of $1\ {\rm AU}\lesssim \tilde{d}\lesssim 
5\ {\rm AU}$.

The Angstrom survey is currently commissioning operation with a real-time 
data processing pipeline and a web-based transient and microlensing alert 
system \citep{darnley07}.  Realization of the alert system and subsequent 
follow-up observations would greatly increase the observation frequency 
and thus overall event rate.  Considering that a significant fraction of 
caustic crossing events are not detected from detections due to the short 
duration between caustic crossings, increasing the monitoring frequency 
from follow-up observations would be able to dramatically increase the 
binary event rate.

Follow-up observation of binary lens events would be important not only 
for the increase of the event rate but also for the characterization of 
lens matter.  The majority of M31 pixel-lensing events are associated 
with giant stars with large angular radii.  The large source size allows 
resolution of caustic crossings because the duration of the caustic 
crossing is proportional to the source radius, i.e.\ 
\begin{equation}
\Delta t_{\rm cc}=2 \left( {\rho_\star \over \sin \alpha}\right)t_{\rm E},
\label{eq11}
\end{equation}
where $\rho_\star=\theta_\star/\theta_{\rm E}$ is the normalized source 
radius and $\alpha$ is the incidence angle of the source trajectory with 
respect to the caustic curve.  For a typical M31 pixel-lensing event with 
$\rho_\star\sim 0.03$ and $t_{\rm E}\sim 10$ days, the caustic-crossing 
duration is $\Delta t_{\rm cc}\gtrsim 14$ hours.  Then, if events can be 
followed up with a frequency  higher than once every hour, the caustic 
crossing can be resolved.  Once the value of $\rho_\star$ is known from 
the analysis of the light curve, one can estimate the Einstein radius 
with additional information of the source radius by $\theta_{\rm E}=
\theta_\star/\rho_\star$.  Since the Einstein radius is related to the 
physical parameters of the lens by the relation in equation~(\ref{eq2}), 
one can better constrain the nature of the lens.

\acknowledgments 
CH and B-GP are supported by the grant (C00072) of the Korea Research 
Foundation.  Y-BJ and C-UL acknowledge the support from Korea Astronomy 
and Space Science Institute.  MI acknowledges the support from the grant 
R01-2006-00-10610-0 provided by the Basic Science Research Program of 
the Korea Science and Engineering Foundation.  EK and JPD are supported, 
respectively, by the Advanced Fellowship and PhD studentship from the UK 
Particle Physics and Astronomy Research Council.  MAI and RGK acknowledge 
the permanent technical and financial support of the Maidanak Observatory 
operations from the consortium of Korea Universities operating under a MOU 
between the consortium and UBAI.

\end{document}